\documentclass[twocolumn,prl,aps]{revtex4}

\def\duzomniejsze{<\kern-.7mm<}
\def\duzowieksze{>\kern-.7mm>}

\def\textbf#1{{\bf #1}}
\def\beq{\begin{equation}}
\def\eeq{\end{equation}}
\def\be{\begin{equation}}
\def\ee{\end{equation}}
\def\ben{\begin{eqnarray}}
\def\een{\end{eqnarray}}
\def\beqa{\begin{eqnarray}}
\def\eeqa{\end{eqnarray}}
\def\eea{\end{array}}
\def\bea{\begin{array}}
\newcommand{\bei}{\begin{itemize}}
\newcommand{\eei}{\end{itemize}}
\newcommand{\bee}{\begin{enumerate}}
\newcommand{\eee}{\end{enumerate}}

\def\hcal{{\cal H}}

\def\>{\rangle}
\def\<{\langle}

\newtheorem{lemma}{Lemma}
\newtheorem{corrolary}{Corrolary}
\newtheorem{theorem}{Theorem}

\begin{document}
\title{Separability of mixed quantum states: linear 
contractions approach}

\author{Micha\l{} Horodecki$^{(1)}$, Pawe\l{} Horodecki$^{(2)}$ 
and Ryszard Horodecki$^{(1)}$}
\affiliation{$^{(1)}$Institute of
Theoretical Physics and Astrophysics, University of Gda\'{n}sk, 80-952
Gda\'{n}sk, Poland}

\affiliation{${}^{(2)}$Faculty of Applied Physics and  Mathematics,
Technical University of Gda\'nsk}

\begin{abstract}
Recently, a new and powerful separability criterion was introduced in 
[O. Rudolph, quant-ph/0202121]  and  [Chen {\it et al.}, quant-ph/0205017].
Composing the main idea behind the above 
criterion and the necessary and sufficient condition in terms of positive maps, 
we provide a characterization of separable states by means of 
linear contractions. The latter need not be positive maps.
We extend the idea to multipartite systems, and find that,
somewhat suprisingly, partial realigment (unlike 
partial transposition) can detect genuinely triparite entanglement.
We also show that for multipartite system any permutation of indices 
of density matrix leads to separability criterion. 
\end{abstract}
\draft
\maketitle

Since the first paper concerning the problem of separability
of mixed states \cite{Werner1989} much effort has been done towards 
operational characterisation of  separable states \cite{Alber2001,Bruss2001-reflections}.
Nevertheless, since 1996 when Peres  designed the 
positive partial transpose (PPT)
test \cite{Peres96}, no better computable separability criterion has been provided 
for a long time. Only quite recently a new,  strong 
criterion for separability (call it {\it realignment criterion}) was 
found by Rudolph  \cite{Rudolph2002-criterion} and  
Chen et al. \cite{Chen2002-criterion}. It is independent of Peres criterion, 
and  turns out to be strong enough to detect entanglement 
of almost all known states, for which the former criterion fails. 
Consequently, it can detect {\it bound entanglement} - a very subtle form 
of entanglement. For some states it is weaker than  PPT one. 

However, as it can be easily checked numerically, the two criteria 
even if taken together cannot detect entanglement of a bound entangled state given in 
\cite{PawelMaciek2000} and \cite{BrussPeres2000} in broad range of parameters. 
So, even though we are now much closer  to operational 
characterisation of entanglement, it is still an open problem. 

There is a necessary and {\it sufficient}  condition of separability \cite{sep1996}
based on positive maps. However, the structure of the set of positive maps 
is still rather unknown, so that the condition is not operational. 
Yet, based on mathematical literature on positive maps, 
it allowed to find several families of entangled state satisfying 
PPT criterion.

The purpose of the paper is double. First we provide
general framework for characterization of 
separability  in terms of linear maps that do not increase trace norm 
on hermitean operators.  Second, we treat multipartite 
states. We extend application of realignment criterion to detect entanglement
of multipartite systems. It is then powerful enough to recognize 
genuinely tripartite entanglement.  Finally we provide a scheme to 
produce separability  criteria via permutations of  matrix indices.

To begin with let us describe how the positive maps characterization 
arose out of Peres criterion.  The core of the 
latter one is that one subjects a subsystem to a positive map -- 
the one that preserve positivity of matrices, i.e. changes density 
matrices again into (possibly unnormalized) density matrixes. 
Consequently, for a positive map $\Lambda$, if the 
total system is in product state $\sigma_A\otimes \sigma_B$,
the resulting operator $(I\otimes \Lambda)\sigma_A\otimes \sigma_B=
\varrho_A\otimes\Lambda(\sigma_B)$ is still a positive operator. 
Then due to linearity also any separable state $\varrho_{AB}=\sum_ip_i
\varrho^i_A\otimes \varrho^i_B$ is mapped into some positive operator.
If we take an initially entangled state, the resulting operator may 
be no longer positive. Thus the negativity of some eigenvalues 
of the resulting operator is indication of entanglement 
of the state. Of course, if the map $I\otimes \Lambda $ 
is positive itself, then it will never  recognize entanglement.
Such maps are called {\it completely positive}. Thus the maps 
that can detect entanglement are the ones that are positive but {\it not}
completely positive.

Let us now describe the result of the present paper.  
One can see that the essential point of realignment criterion 
from \cite{Rudolph2002-criterion,Chen2002-criterion}
is  some linear map $L$ (acting on both subsystems) that does not increase 
trace norm of product states $||L(\sigma_A \otimes \sigma_B)|| \leq 1$.  
Then by convexity of norm, also for separable states we will have 
$||L(\varrho^{sep}_{AB})||\leq 1$.    Thus {\it any linear map that does 
not increase norm of product states constitute separability 
necessary condition for separability}.  To be nontrivial, 
at least for some state it should happen 
that $||L(\varrho_{AB})||>1$, and this is indication 
of entanglement.

The role of $L$ in Refs. 
\cite{Rudolph2002-criterion,Chen2002-criterion}
was played by the realigment map $L^{r}$ acting 
on product operators as follows: $L^{r}(A \otimes B)=|A\rangle \langle B^*|$,
where the latter involves bra-ket formalism of Hilbert Schmidt 
space of operators with scalar product 
$\langle A|B\rangle=Tr(A^{\dagger}B)$.
The above map was previously discussed in context 
of general matrix algebra \cite{Oxenrider85}, completely 
positive and completely copositive  maps \cite{Yopp00}
and positive maps theory \cite{Karol2001}.

To make better analogy with our previous condition,
consider $L$ acting only on subsystem. If $L$ is contraction in 
trace norm, i.e. if it does not increase the trace norm, 
then on product states we have $||(I\otimes L)(\sigma_A\otimes \sigma_B)||
= ||\sigma_A||\, ||L(\sigma_B)||\leq 1$. Thus instead of 
positive map $\Lambda$ we have contraction $L$ and we check 
trace norm instead of positivity. Again, if a map $I\otimes L$ 
is itself contraction, then it will be useless: simply all states 
will satisfy the criterion. Thus, the relevant maps are 
the contractions that are not {\it complete contractions}, i.e. 
$I\otimes L$ is no longer  a contraction. It should be noted here that 
to obtain  sufficient condition, we will need the set of maps 
that are contractions on hermitean operators.

Example of the map  
being {\it contraction but not complete contraction} 
is  transposition $T$ (which is at the same time 
positive map). $I\otimes T$ acting on maximally entangled state 
of $d\otimes d$ system results in operator of norm $d$ 
(to deal with such oddities of trace norm, one designed so called 
completely bounded norm, see \cite{WernerHolevo1999-bosonic}). 
Below we will show that the set of such criteria constitutes 
necessary and sufficient condition. However, the map should 
have output dimension twice as much as the dimension of 
the subsystem. 

To begin with, let us recall the positive maps condition \cite{sep1996}. 
\begin{theorem}
The following statements are equivalent.
\bee
\item A state $\varrho$ acting on Hilbert space $C^d\otimes C^d$ is 
separable 
\item For any  positive map 
$\Lambda: M_d\to M_{d}$
\be
(I\otimes \Lambda)\varrho_{AB}\geq 0
\ee
\item 
For any  operator $W\in M_d\otimes M_d$ which satisfies 
\be
Tr(W \sigma^{sep}_{AB})\geq 0
\label{eq-pos-prod}
\ee
for all separable  states $\sigma^{sep}_{AB}$ 
we have 
\be
Tr W \varrho_{AB}\geq 0
\ee
\eee 
\end{theorem}

{\bf Remark.} The operators $W$ of item (3)
which are non-positive give nontrivial necessary conditions 
for separability  and are called {\it entanglement witnesses} 
(see e.g. \cite{TerhalBell2000}).
Slightly abusing terminology, we will call $W$ entanglement witnesses,
even if we do not know if it is non-positive operator.

To pass from positive maps to linear  contractions
we need to show that it suffices to restrict to trace-preserving 
maps.  
\begin{lemma}
A state $\varrho$ acting on Hilbert space 
$\hcal_A\otimes \hcal_B=C^d\otimes C^d$ is 
separable iff for any trace preserving positive map 
$\Lambda: M_d\to M_{2d}$
\be
(I\otimes \Lambda)\varrho_{AB}\geq 0
\ee
\end{lemma}

{\bf Remark.} Note that we need larger dimension of output of 
$\Lambda$ than in the condition 2 of Theorem 1 where we do not require 
the map to be trace preserving.
It may be worth to mention that there 
is a dual proposition \cite{Pawel2001-NATO}  stating that 
bipartite states are completely characterised by identity preserving 
positive maps which does not require such output dimesion enlargement.
  
{\bf Proof.}
It suffices to show that for any entangled state we will find a 
positive trace presering map, such that operator 
$I\otimes \Lambda(\varrho_{AB})$ is not positive. 
We will follow techniques used in \cite{Pawel2001-NATO}. 
Let $\varrho_{AB}$ be entangled. Then there exists entanglement witness $W$  
such that 
\be
Tr(W \varrho_{AB})<0.
\ee 
Consider operator $W_B$  being  reduction of $W$ (one can see that even though $W$ is not a
positive opreator, its reduction is). Let $P_B$ be the projector onto support 
of $W_B$ and $P_B^\perp$ the complementary one, so that $P_B+P^\perp_B=I_B$.
We then introduce another  witness $W'$  given by
\be
W'=W+I_A \otimes  P_B^\perp.
\ee
One finds that  $W'$ satisfies the condition (\ref{eq-pos-prod}).  It satisfies  also 
\be
Tr(W' I_A\otimes  \hat P_B \varrho_{AB})<0 
\ee
where we used notation 
$I \otimes \hat X\varrho_{AB}= I \otimes X \varrho_{AB} I \otimes X^\dagger $.
Its   reduction $W'_B$ is  now of full rank so that we can introduce a new witness 
$W''=I \otimes {\sqrt{W_B'}^{-1}} \, W'\, I \otimes {\sqrt{W_B'}^{-1}}$. 
The witness $W''$ has reduction
$W''_B$ equal to identity and satisfies
\be
Tr(W'' I  \otimes  \hat V\varrho_{AB})<0
\ee
where $V=\sqrt{W_B'} P_B$.
Now we take positive map $\Gamma$ satisfying 
$(I\otimes \Gamma ) d P_+=W''$ \cite{Jamiolkowski}, where $P_+=|\psi_+\>\<\psi_+|$ 
with $\psi_+={1\over \sqrt d} \sum_i |i\> \otimes |i\>$.  This implies 
\be
Tr (P_+ (I\otimes \Gamma^\dagger \hat V\ )\varrho_{AB})<0
\ee
Since $W''_B=I$, the map $\Gamma$ is 
identity preserving, so that its dual map $\Gamma^\dagger $ 
is trace preserving. We have not yet obtained the thesis of 
the theorem, as there is still the operator  $V$ involved. 
We have thus proved that for any entangled state $\varrho$ 
there exists a trace preserving map $\Gamma$ and an operator 
(local filter \cite{Gisin1996-filters,BBPS1996}) $V$ such that 
$( I \otimes \Gamma \hat V)\varrho_{AB}$ is not positive. 
We build the needed map $\Lambda$ as follows. We add a  qubit 
on Bob's side, and consider the following CP map
\be
\Lambda_{cp} (\cdot) =\hat  V (\cdot) \otimes |0\>\<0|
+\hat V' (\cdot) \otimes |1\>\<1|
\ee
where $V'$ is so chosen that $V'^\dagger V'+ V^\dagger V=I$. 
The map is thus simply generalized measurement on 
Bob's system with writing record on the qubit. 
After the map $\Lambda_{cp}$ 
we perform the map $\Gamma$. Since after the first map the state 
is separable against Bob's systems and ancillar qubit,
the composition $\Lambda =\Gamma \Lambda_{cp}$ is positive map. 
Moreover we have 
$Tr \bigl[ ((I\otimes \Lambda)\varrho_{AB}) (P_+\otimes |0\>\<0|)\bigr]<0$.
The new map is composition of trace-preserving maps, so that it is 
itself trace preserving.  Thus we have found the needed map for 
arbitrary entangled state, which ends the proof.

Now, to pass to maps that do not need to be positive any longer, 
we need to  formulate of the above condition by means of trace norm.
It is possible, because we have restricted to trace preserving maps. 
In result the operator $(I\otimes \Lambda)\varrho_{AB}$ has unit trace,
and its positivity is equivalent to having unit trace norm. Thus we have 
obtained 

\begin{corrolary}
\label{th-warunek1}
A state $\varrho$ acting on Hilbert space $C^d\otimes C^d$ is 
separable iff for any trace preserving positive map 
$\Lambda: M_d\to M_{2d}$
\be
||(I\otimes \Lambda)\varrho_{AB}||\leq 1 
\ee
\end{corrolary}
Instead of equality $||(I\otimes \Lambda)\varrho_{AB}||=1 $
we put inequality, which is irrelevant for the requirement, but 
allows us to give up the positivity condition. Namely we have 

\begin{theorem}
A state $\varrho_{AB}$ acting on Hilbert space $C^d\otimes C^d$ is 
separable iff for any 
linear map $L: M_d\to M_{2d}$ which is contraction on hermitean operators 
one has
\be
||(I\otimes L)\varrho_{AB}||\leq 1 
\ee
\end{theorem}

{\bf Remark.} Instead of contraction on hermitean operators, 
we could take contractions on positive operators (or just on projectors).

{\bf Proof.}
The sufficiency is obvious, as the condition is certainly no weaker 
than that of lemma \ref{th-warunek1}. Indeed it is well known 
that a positive trace preserving map does not increase trace norm 
on hermitean operators.
To show the necessity we note as in Ref. \cite{Chen2002-criterion}. 
Due to convexity of  norm, it suffices  to consider only product 
states $\sigma_A\otimes \sigma_B$.  Since $||\sigma_A||=||\sigma_B||=1$ 
we obtain 
\be
||(I\otimes L)(\sigma_A\otimes \sigma_B)||=||\sigma_A|| 
||L(\sigma_B)||\leq 1. 
\ee
where we used the fact that $L$ is contraction hermitean  (hence 
also positive)  operators. This ends the proof.

The condition has similar structure as the conditions based 
on positive maps. In the latter one, the key role was played by the maps 
that are positive but are not completely positive i.e. 
$\Lambda$ is positive, but $I\otimes \Lambda$ is not. Here the 
analogous role is played  by the contractions 
that are not completely-contracting, i.e. such that $L$ is 
contraction, but  $I\otimes L$ is not. Example is 
transposition (see, e.g.   \cite{WernerHolevo1999-bosonic}). 

One can also express the characterization of separable states 
by means of  linear maps that are 
contractions on product states. The necessary and sufficient 
condition then has the form: {\it the state $\varrho_{AB}$ is separable 
iff for any linear map $L$ satisfying
$||L(\sigma_A\otimes \sigma_B)||\leq 1$ for all states 
$\sigma_A$, $\sigma_B$ one has 
\be
||L(\varrho_{AB})||\leq 1. 
\ee
}
Example of $L$ can be product of contractions, but $L$ need not be 
product this time. Example of nonproduct map is just the 
map $L^r$ of Refs. \cite{Rudolph2002-criterion,Chen2002-criterion}
we recalled in introduction.

Let us now pass to multipartite  systems. One can imediately see that 
the method can be applied to  this case. In  Ref. \cite{multisep} we 
have provided characterisation  of entanglement of multipartite 
states which employed positive maps  on product states.
Here we will employ linear maps that are contractions on product states.
Consider $n$-partite system.  Thus any given linear map $L$ 
acting on $k\leq n$, which is contraction on product states $P_1\otimes \ldots\otimes P_k$ 
gives rise to the following separability criterion: if a 
state $\varrho$ is separable,
then 
\begin{equation}
||L_{(k)}\otimes I_{(n-k)}\varrho||\leq 1 
\label{multi}
\end{equation}
 where the indices mean that 
we act by $L$ on chosen $k$ subsystems, while leave untouched 
the remaining $n-k$ subsystems. We are not able to prove here 
that criteria of this type constitute sufficient condition.  However,
strong criteria can be obtained in this way. 

To  show the power of such approach we recall  that due to a richer structure,
the  multipartite  systems can exhibit a peculiar phenomenon: there 
are tripartite states  that are separable under any 
bipartite cut ({\it biseparable}), but  nevertheless are entangled
(are not mixtures of fully product states $|\psi_1\>|\psi_2\>|\psi_3\>$).
They contain genuinely tripartite entanglement, and must be necessary mixed:
pure states cannot share such feature. Such entanglement cannot be 
detected by  methods involving linear map acting on one subsystem. 
Indeed, such a map will detect only the bipartite entanglement between 
the  subsystem and the cluster of all other subsystems. Partial transpose
even if applied to several systems can also detect only bipartite  
entanglement. The realignement  map  $L^r$ acts however on two systems 
and it is a contraction on product states, 
hence it is  a good candidate to detect genuinely tripartite  entanglement.
First example of entangled biseparable state was given in 
Ref. \cite{BennettUPBI1999} where also it was pointed out that 
it represent so called bound entanglement.
The example is a three-qubit state of the form 
\be
{\tilde\varrho}_{ABC}={1\over 8} (I-\sum_{i=1}^4|\psi_i\>\<\psi_i|)
\label{bip}
\ee
where $\psi_i$'s are given by 
\be
|0,1,+\>,|1,+,0\>,|+,1,0\>,|-,-,-\>
\ee
with $|\pm\>={1/\sqrt2}(|0\>\pm|1\>)$. 
The above state is called biseparable 
(or semiseparable) because it is separable under any bipartite cut 
$A|BC$, $B|CA$, $C|AB$. Still it is entangled. 
Apart form, in general not easy, range analysis (see 
\cite{BennettUPBI1999}) basing on tripartite version 
of range separability criterion \cite{Pawel97} 
there was no method to detect its entanglement. 
Here we report fully operational criterion doing that:
it is a particular form of (\ref{multi})
namely application of the realignment map to any of two qubits.
Indeed we found numerically 
that the norm $||[I_{A} \otimes 
L^{r}_{BC}](\tilde{\varrho}_{ABC})||$ 
amounts to $1.08649$.

Finally, we will exhibit a general method of producing 
separability criteria for multipartite systems. 
For bipartite  systems it will produce only Peres criterion 
and the realignment one. For multipartite systems we leave the problem 
of equivalence of criteria open, though we hope that many of 
them will be non-equivalent to each other. We will explain 
the idea on bipartite systems, since it 
immediately generalizes to general case.
The idea is to take linear map as any permutation $\tau$ of 
density matrix indices, if written in a  product basis 
(such permutations were considered recently in the context of separability 
in Ref. \cite{Karol2001}). 
Such a linear map can be represented by its action on basic operators 
of the form $|i\>\<j|\otimes |k\>\<l|$:
\be
L(|i\>\<j|\otimes |k\>\<l|)= |\tau(i)\>\<\tau(j)|\otimes |
\tau(k)\>\<\tau(l)|
\ee 
To show that $L$ gives rise to the criterion, it suffices to note that 
it keeps norm for pure product states. Indeed the latter 
are belong to the set of operators of  the general 
form $|\psi_1\>\<\psi_2|\otimes |\psi_3\>\<\psi_4|$. $L$ transforms 
such operators  into 
$|\psi_{\tau(1)}^{\#}\>\<\psi_{\tau(2)}^{\#}|\otimes 
|\psi_{\tau(3)}^{\#}\>
\<\psi_{\tau(4)}^{\#}|$, where $\psi^\#_i=\psi^*_i$ 
if $\tau(i)$ has a different parity than $i$ 
and $\psi^\#_i=\psi_i$ otherwise. 
Thus the output is of the same general form as the input,
which, in turn has unit norm, due to the fact that 
trace norm $||X||$ does not change under transformation 
$X\to U X V$ where $U,V $ 
are unitary.

For example partial transpose (on right subsystem) coresponds to 
permutation $\tau=(34)$. This gives  $|i\>\<j|\otimes |k\>\<l|\to
|i\>\<j|\otimes |l\>\<k|$, and, consequently 
$|\psi\>\<\psi|\otimes |\phi\>\<\phi|\to|\psi\>\<\psi|
\otimes |\phi^*\>\<\phi^*|$.
Realignment corresponds to  permutation  $\tau=[1234\to 2413]$ 
which produces  transformation $|\psi\>\<\psi|\otimes |\phi\>\<\phi|\to
|\psi^*\>\<\phi|\otimes |\psi\>\<\phi^*|$.
Using the fact that trace norm is invariant under full transpose
and  the map $X\to U X V$ one can find that 
all permutations are equivalent to either realignement
or partial transpose or identity (which does not, of course, produce separability criterion).
For example a permutation $\tau=(23)$ gives rise to a map 
which is equivalent to original realignment.

It is obvious that the method works also for multipartite case.
For tripartite systems one can find that nontrivial permutations 
can be chosen as $[12**** \to 1*2***]$ or $[12****\to 12****]$. The 
notation has to be understood as follows:
in the first case $\tau(1)=1$ and $\tau(2)=3$ 
and in the second case 
$\tau(1)=1$ and $\tau(2)=2$. It is yet not  clear which 
of them are nonequivalent. To have some function that would 
indicate joint power of the criteria, one can take 
$E=\sup_\tau||\varrho_{\tau({\mathbf i})}||$ 
where ${\mathbf i}$ stands for multiindex denoting all indices of $\varrho$ 
in product basis. It remains open question whether $E$ can be good 
entanglement measure, though it definitely indicates all entanglement 
that can be detected by permutation criteria.

To conclude, we were able to extract a general structure behind the 
realignment criterion. We then provided framework that 
incorporates the latter one as well as partial transpose.
As a result  we provided  a method to produce a whole bunch of separability 
criteria for multipartite systems. Example of such criterion
is realignment applied to two subsystems of triparite systems. 
It is powerful enough to detect a genuinely tripartite entanglement.

We would like to thank Robert Alicki, Piotr Badzi\c a{}g, 
Karol Horodecki and  Karol \.Zyczkowski for useful discussions.  

\bibliographystyle{apsrev}


\bibliography{ref} 

\begin{thebibliography}{21}
\expandafter\ifx\csname natexlab\endcsname\relax\def\natexlab#1{#1}\fi
\expandafter\ifx\csname bibnamefont\endcsname\relax
  \def\bibnamefont#1{#1}\fi
\expandafter\ifx\csname bibfnamefont\endcsname\relax
  \def\bibfnamefont#1{#1}\fi
\expandafter\ifx\csname citenamefont\endcsname\relax
  \def\citenamefont#1{#1}\fi
\expandafter\ifx\csname url\endcsname\relax
  \def\url#1{\texttt{#1}}\fi
\expandafter\ifx\csname urlprefix\endcsname\relax\def\urlprefix{URL }\fi
\providecommand{\bibinfo}[2]{#2}
\providecommand{\eprint}[2][]{\url{#2}}

\bibitem[{\citenamefont{Werner}(1989)}]{Werner1989}
\bibinfo{author}{\bibfnamefont{R.}~\bibnamefont{Werner}},
  \bibinfo{journal}{Phys. Rev.} \textbf{\bibinfo{volume}{A 40}},
  \bibinfo{pages}{4277} (\bibinfo{year}{1989}).

\bibitem[{\citenamefont{Alber et~al.}(2001)\citenamefont{Alber, Beth,
  Horodecki, Horodecki, Horodecki, Rotteler, Weinfurter, Werner, and
  Zeilinger}}]{Alber2001}
\bibinfo{author}{\bibfnamefont{G.}~\bibnamefont{Alber}},
  \bibinfo{author}{\bibfnamefont{T.}~\bibnamefont{Beth}},
  \bibinfo{author}{\bibfnamefont{M.}~\bibnamefont{Horodecki}},
  \bibinfo{author}{\bibfnamefont{P.}~\bibnamefont{Horodecki}},
  \bibinfo{author}{\bibfnamefont{R.}~\bibnamefont{Horodecki}},
  \bibinfo{author}{\bibfnamefont{M.}~\bibnamefont{Rotteler}},
  \bibinfo{author}{\bibfnamefont{H.}~\bibnamefont{Weinfurter}},
  \bibinfo{author}{\bibfnamefont{R.}~\bibnamefont{Werner}}, \bibnamefont{and}
  \bibinfo{author}{\bibfnamefont{A.}~\bibnamefont{Zeilinger}},
  \emph{\bibinfo{title}{Quantum Information: An Introduction to Basic
  Theoretical Concepts and Experiments}} (\bibinfo{publisher}{Springer},
  \bibinfo{year}{2001}).

\bibitem[{\citenamefont{Bruss et~al.}()\citenamefont{Bruss, Cirac, Horodecki,
  Hulpke, Kraus, Lewenstein, and Sanpera}}]{Bruss2001-reflections}
\bibinfo{author}{\bibfnamefont{D.}~\bibnamefont{Bruss}},
  \bibinfo{author}{\bibfnamefont{J.~I.} \bibnamefont{Cirac}},
  \bibinfo{author}{\bibfnamefont{P.}~\bibnamefont{Horodecki}},
  \bibinfo{author}{\bibfnamefont{F.}~\bibnamefont{Hulpke}},
  \bibinfo{author}{\bibfnamefont{B.}~\bibnamefont{Kraus}},
  \bibinfo{author}{\bibfnamefont{M.}~\bibnamefont{Lewenstein}},
  \bibnamefont{and} \bibinfo{author}{\bibfnamefont{A.}~\bibnamefont{Sanpera}},
  \bibinfo{note}{(to appear in J. Mod. Opt.)}, \eprint{quant-ph/0110081}.

\bibitem[{\citenamefont{Peres}(1996)}]{Peres96}
\bibinfo{author}{\bibfnamefont{A.}~\bibnamefont{Peres}},
  \bibinfo{journal}{Phys. Rev. Lett.} \textbf{\bibinfo{volume}{76}},
  \bibinfo{pages}{1413} (\bibinfo{year}{1996}).

\bibitem[{\citenamefont{Rudolph}()}]{Rudolph2002-criterion}
\bibinfo{author}{\bibfnamefont{O.}~\bibnamefont{Rudolph}},
  \eprint{quant-ph/0202121}.

\bibitem[{\citenamefont{Chen et~al.}()\citenamefont{Chen, Wu, and
  Yang}}]{Chen2002-criterion}
\bibinfo{author}{\bibfnamefont{K.}~\bibnamefont{Chen}},
  \bibinfo{author}{\bibfnamefont{L.-A.} \bibnamefont{Wu}}, \bibnamefont{and}
  \bibinfo{author}{\bibfnamefont{L.}~\bibnamefont{Yang}},
  \eprint{quant-ph/0205017}.

\bibitem[{\citenamefont{Horodecki and Lewenstein}(2000)}]{PawelMaciek2000}
\bibinfo{author}{\bibfnamefont{P.}~\bibnamefont{Horodecki}} \bibnamefont{and}
  \bibinfo{author}{\bibfnamefont{M.}~\bibnamefont{Lewenstein}},
  \bibinfo{journal}{Phys. Rev. Lett} \textbf{\bibinfo{volume}{85}},
  \bibinfo{pages}{2657} (\bibinfo{year}{2000}), \eprint{quant-ph/0001035}.

\bibitem[{\citenamefont{Bruss and Peres}(2000)}]{BrussPeres2000}
\bibinfo{author}{\bibfnamefont{D.}~\bibnamefont{Bruss}} \bibnamefont{and}
  \bibinfo{author}{\bibfnamefont{A.}~\bibnamefont{Peres}},
  \bibinfo{journal}{Phys. Rev. A} \textbf{\bibinfo{volume}{61}},
  \bibinfo{pages}{30301(R)} (\bibinfo{year}{2000}), \eprint{quant-ph/9911056}.

\bibitem[{\citenamefont{Horodecki et~al.}(1996)\citenamefont{Horodecki,
  Horodecki, and Horodecki}}]{sep1996}
\bibinfo{author}{\bibfnamefont{M.}~\bibnamefont{Horodecki}},
  \bibinfo{author}{\bibfnamefont{P.}~\bibnamefont{Horodecki}},
  \bibnamefont{and}
  \bibinfo{author}{\bibfnamefont{R.}~\bibnamefont{Horodecki}},
  \bibinfo{journal}{Phys. Lett} \textbf{\bibinfo{volume}{223}},
  \bibinfo{pages}{1} (\bibinfo{year}{1996}), \eprint{quant-ph/9605038}.

\bibitem[{\citenamefont{Oxenrider and Hill}(1985)}]{Oxenrider85}
\bibinfo{author}{\bibfnamefont{C.~J.} \bibnamefont{Oxenrider}}
  \bibnamefont{and} \bibinfo{author}{\bibfnamefont{R.~D.} \bibnamefont{Hill}},
  \bibinfo{journal}{Lin. Alg. Appl.} \textbf{\bibinfo{volume}{69}},
  \bibinfo{pages}{205} (\bibinfo{year}{1985}).

\bibitem[{\citenamefont{Yopp and Hill}(2000)}]{Yopp00}
\bibinfo{author}{\bibfnamefont{D.~A.} \bibnamefont{Yopp}} \bibnamefont{and}
  \bibinfo{author}{\bibfnamefont{R.~D.} \bibnamefont{Hill}},
  \bibinfo{journal}{Lin. Alg. Appl.} \textbf{\bibinfo{volume}{312}},
  \bibinfo{pages}{1} (\bibinfo{year}{2000}).

\bibitem[{\citenamefont{\.Zyczkowski}()}]{Karol2001}
\bibinfo{author}{\bibfnamefont{K.}~\bibnamefont{\.Zyczkowski}},
  \bibinfo{note}{preprint 2001, (unpublished)}.

\bibitem[{\citenamefont{Werner and Holevo}()}]{WernerHolevo1999-bosonic}
\bibinfo{author}{\bibfnamefont{R.~F.} \bibnamefont{Werner}} \bibnamefont{and}
  \bibinfo{author}{\bibfnamefont{A.~S.} \bibnamefont{Holevo}},
  \eprint{quant-ph/9912067}.

\bibitem[{\citenamefont{Terhal}(2000)}]{TerhalBell2000}
\bibinfo{author}{\bibfnamefont{B.~M.} \bibnamefont{Terhal}},
  \bibinfo{journal}{Phys. Lett. A.} \textbf{\bibinfo{volume}{271}},
  \bibinfo{pages}{319} (\bibinfo{year}{2000}), \eprint{quant-ph/9911057}.

\bibitem[{\citenamefont{Horodecki}(2001)}]{Pawel2001-NATO}
\bibinfo{author}{\bibfnamefont{P.}~\bibnamefont{Horodecki}}, in
  \emph{\bibinfo{booktitle}{Proceedings of NATO ARW:``Decoherence and its
  Implications in Quantum Computation and Information Transfer''}}, edited by
  \bibinfo{editor}{\bibfnamefont{T.}~\bibnamefont{Gonis}} \bibnamefont{and}
  \bibinfo{editor}{\bibfnamefont{P.~E.~A.} \bibnamefont{Turchi}}
  (\bibinfo{publisher}{IOS press}, \bibinfo{year}{2001}), Series III - Comp.
  and Sci. Sys., p. \bibinfo{pages}{299}.

\bibitem[{\citenamefont{Jamio\l{}kowski}(1972)}]{Jamiolkowski}
\bibinfo{author}{\bibfnamefont{A.}~\bibnamefont{Jamio\l{}kowski}},
  \bibinfo{journal}{Rep. Math. Phys.} \textbf{\bibinfo{volume}{3}},
  \bibinfo{pages}{275} (\bibinfo{year}{1972}).

\bibitem[{\citenamefont{Bennett et~al.}(1996)\citenamefont{Bennett, Bernstein,
  Popescu, and Schumacher}}]{BBPS1996}
\bibinfo{author}{\bibfnamefont{C.~H.} \bibnamefont{Bennett}},
  \bibinfo{author}{\bibfnamefont{H.}~\bibnamefont{Bernstein}},
  \bibinfo{author}{\bibfnamefont{S.}~\bibnamefont{Popescu}}, \bibnamefont{and}
  \bibinfo{author}{\bibfnamefont{B.}~\bibnamefont{Schumacher}},
  \bibinfo{journal}{Phys. Rev. A} \textbf{\bibinfo{volume}{53}},
  \bibinfo{pages}{2046} (\bibinfo{year}{1996}), \eprint{quant-ph/9511030}.

\bibitem[{\citenamefont{Gisin}(1996)}]{Gisin1996-filters}
\bibinfo{author}{\bibfnamefont{N.}~\bibnamefont{Gisin}},
  \bibinfo{journal}{Phys. Lett. A} \textbf{\bibinfo{volume}{210}},
  \bibinfo{pages}{151} (\bibinfo{year}{1996}).

\bibitem[{\citenamefont{Horodecki et~al.}(2001)\citenamefont{Horodecki,
  Horodecki, and Horodecki}}]{multisep}
\bibinfo{author}{\bibfnamefont{M.}~\bibnamefont{Horodecki}},
  \bibinfo{author}{\bibfnamefont{P.}~\bibnamefont{Horodecki}},
  \bibnamefont{and}
  \bibinfo{author}{\bibfnamefont{R.}~\bibnamefont{Horodecki}},
  \bibinfo{journal}{Phys. Lett. A} \textbf{\bibinfo{volume}{283}},
  \bibinfo{pages}{1} (\bibinfo{year}{2001}), \eprint{quant-ph/0006071}.

\bibitem[{\citenamefont{Bennett et~al.}(1999)\citenamefont{Bennett, DiVincenzo,
  Mor, Shor, Smolin, and Terhal}}]{BennettUPBI1999}
\bibinfo{author}{\bibfnamefont{C.~H.} \bibnamefont{Bennett}},
  \bibinfo{author}{\bibfnamefont{D.~P.} \bibnamefont{DiVincenzo}},
  \bibinfo{author}{\bibfnamefont{T.}~\bibnamefont{Mor}},
  \bibinfo{author}{\bibfnamefont{P.~W.} \bibnamefont{Shor}},
  \bibinfo{author}{\bibfnamefont{J.~A.} \bibnamefont{Smolin}},
  \bibnamefont{and} \bibinfo{author}{\bibfnamefont{B.~M.}
  \bibnamefont{Terhal}}, \bibinfo{journal}{Phys. Rev. Lett.}
  \textbf{\bibinfo{volume}{53}}, \bibinfo{pages}{5385} (\bibinfo{year}{1999}),
  \eprint{quant-ph/9808030}.

\bibitem[{\citenamefont{Horodecki}(1997)}]{Pawel97}
\bibinfo{author}{\bibfnamefont{P.}~\bibnamefont{Horodecki}},
  \bibinfo{journal}{Phys. Lett. A} \textbf{\bibinfo{volume}{232}},
  \bibinfo{pages}{333} (\bibinfo{year}{1997}).

\end{thebibliography}
\end{document}